\documentclass{article}
\newcommand{\be}{\begin{equation}}
\newcommand{\ee}{\end{equation}}

\def\bea{\begin{eqnarray}}
\def\eea{\end{eqnarray}}
\def\bean{\begin{eqnarray*}}
\def\eean{\end{eqnarray*}}

\def\part#1{\partial^{}_#1}

\newcommand{\barr}{\begin{array}}
\newcommand{\earr}{\end{array}}

\newcommand{\noi}{\noindent}
\newcommand{\nnum}{\nonumber}

\newcommand{\bed}{\begin{displaymath}}
\newcommand{\eed}{\end{displaymath}}
\newcommand{\bal}{\begin{array}{ll}}
\newcommand{\eal}{\end{array}}

\newcommand{\cq}{{\cal Q}}

\newcommand{\tgamma}{\tilde{\gamma}}
\newcommand{\csr}{{}_{\rm CSR}}

\begin{document}
\begin{titlepage}
\begin{flushright} UFIFT-HEP-02-16 \\
\end{flushright}
\vskip .5cm
\centerline{\LARGE{\bf {Continuous Spin Representations of the}}}
\vskip .2cm
\centerline{\LARGE{\bf  Poincar\'e and Super-Poincar\'e Groups }}
\vskip 1cm
\date{\today}
\centerline{\bf Lars Brink }
\vskip .5cm
\centerline{\em Department of Theoretical Physics}\centerline{\em Chalmers
University of Technology,}
\centerline{\em   S-412 96 G\"oteborg, Sweden}
\vskip 1cm
\centerline{\bf Abu M. Khan, Pierre Ramond, Xiaozhen Xiong} 
\vskip .5cm
\centerline{\em  Institute for Fundamental Theory,}
\centerline{\em Department of Physics, University of Florida}
\centerline{\em Gainesville FL 32611, USA}
\vskip 1.5cm
\centerline{\bf {Abstract}}
\vskip .5cm \noindent
We construct Wigner's continuous spin representations of the Poincar\'e algebra
 for massless particles in higher dimensions. 
The states are labeled both by the length of a space-like  
translation vector and the Dynkin indices of the {\it short little 
group} $SO(d-3)$, where $d$ is the space-time dimension.  Continuous spin representations 
are in one-to-one correspondence with representations of the short little group. 
We also demonstrate how combinations of  the bosonic and fermionic representations form supermultiplets of the super-Poincar\'e algebra.  
If the light-cone translations are nilpotent, these representations 
become finite dimensional, but  contain zero or negative norm 
states, and their supersymmetry algebra contains a central charge in 
four dimensions.
\vfill
\end{titlepage}
\section{Introduction}
The   Poincar\'e group  is an essential ingredient of relativistic quantum
field theories. Its representations in four space-time dimensions were
first studied by E. Wigner~\cite{WIGNER}. Some of its representations describe 
quantum states found in local field theory: massless particles of 
fixed helicity and massive particles with or without spin. In string theory, 
the ``front form" construction of the Poincar\'e group generators~\cite{DIRAC, BC} 
has been particularly useful in describing the $26$-dimensional bosonic string~\cite{GGRT}. 

Other representations do not seem to
be realized by Nature. One is the tachyon representation with negative
mass-squared which appears in theories as indicator  of instabilities,
for instance in spontaneous symmetry breaking. The others are the
``continuous spin representations" (CSR), which describe a massless object with
an infinity number of helicities. Wigner himself has argued~\cite{WIGNER2}
against their use in Physics since they lead to infinite heat capacity of
the vacuum.  
All attempts to associate these representations with
physical systems have failed. Several authors~\cite{ABBOTT,HIRATA} have
shown their naive quantization implies either  non-locality or a
breakdown of causality.  More recently, Zoller~\cite{ZOLLER} showed
how they could arise from a higher-derivative Lagrangian such as the
length of the acceleration. 
In addition, 
massless theories with helicities greater than two have special problems~\cite{WITTENWEINBERG,DESER}, unless perhaps  for theories with an infinite number of degrees of
freedom.  Not surprisingly, these representations have been forgotten.

Yet, the idea of an infinite number of massless
spins has some attractions. Indeed, Vasiliev~\cite{VASILIEV} among others has argued for the
existence of such theories. A particular lure is the zero tension
(infinite slope) limit of string theories which suggests (classically) an
infinite number of massless states with unbounded helicities.

The object of this paper is to study these representations in
higher dimensions, and show that they can be assembled in representations of supersymmetry.

\section{The Poincar\'e Algebra}
The Poincar\'e generators  satisfy the commutation
relations,
\bea \nonumber
[P^\mu , P^\nu]	& = & 0 \ , \\ \nonumber
 [ M^{\mu \nu} , P^\sigma ] & =	& i (\eta^{\mu \sigma} P^\nu -
\eta^{\nu
\sigma} P^\mu)\ , \\ \nonumber
[ M^{\mu \nu} , M^{\alpha \beta} ] & = & i (\eta^{\mu \alpha} M^{\nu
\beta} + \eta^{\alpha \nu} M^{\beta \mu} + \eta^{\nu \beta} M^{\mu \alpha}
+ \eta^{\beta \mu} M^{\alpha \nu}) \ ,
\eea
where $\eta_{\mu\nu}=(-1,1,\cdots,1)$. Its representations are
characterized by the values of the  Casimir operators which are the squared mass $P_\mu P^\mu$, and
the squares of the Pauli-Lubanski forms built out of the Levi-Civit\'a
symbols. 
In $d$ space-time dimensions, the Pauli-Lubanski $n$-forms are,
\be \label{forms}
W_{\mu_1 \cdots \mu_n} ~=~ \frac{\epsilon_{\mu_1 \cdots \mu_n \mu_{n+1} 
\cdots \mu_d} P^{\mu_d^{}}
M^{\mu_{n+1}\mu_{n+2}} \cdots M^{\mu_{d-2}^{} \mu_{d-1}^{}}}
{{\sqrt{n! \ 2_{}^{\frac{d-n+1}{2}}  \;\left(\frac{d-n-1}{2}\right)\, 
\left(\frac{d-n-1}{2}\right)!}} } \ , 
\ee
where $n=1,3, \cdots , (d-3)$ for $d$ even and $n=0,2,\cdots , (d-3)$ for
$d$ odd.

It is convenient to express the  Poincar\'e generators in
Dirac's~\cite{DIRAC} front form, derived long ago for spin in four dimensions by Bacry and 
Chang~\cite{BC}. For a particle of
mass $m$, the Poincar\'e  generators are expressed in terms of transverse
positions and momenta
\be
[\,x^i~\,,\,~p^j\,]~=~i\,\delta^{ij}\ ,~i,j=1,2,...,d-2\ ;\qquad
[\,x^-\,,\,p^+\,]~=~-i\ ,\ee
and
\be
x^{\pm}~=~\frac{1}{\sqrt{2}}(x^0\pm x^{d})\ ,\qquad
p^{\pm}~=~\frac{1}{\sqrt{2}}(p^0\pm p^{d})\ .
\ee
The  translations  are given  by
\be
P_{}^-~=~\frac{p^i_{}p^i_{}+m^2_{}}{2p_{}^+}\ ,\qquad P_{}^+~=~p_{}^+\
,\qquad P_{}^i~=~p_{}^i\ ,\ee
where $P^-$ is the light-cone Hamiltonian. The Lorentz
generators break up into those which  transform the transverse plane into
itself, and those which
transform out of that plane, (called ``kinematic" and ``Hamiltonian",
respectively by Dirac). The kinematic generators are given by
\be
M_{}^{+i}~=~-x_{}^ip_{}^+\ ,\qquad M_{}^{+-}~=~-x_{}^-p_{}^+\ ,\qquad
M^{ij}~=~x^ip^j-x^jp^i+S^{ij}_{}\ ,
\ee
where $S^{ij}_{}$ obey the $SO(d-2)$ Lie algebra of the transverse little
group
\be
[\,S^{ij}_{}\,,\,S^{kl}_{}\,]~=~i(\delta^{ik}_{}\,S^{jl}_{}+\delta^{jl}_{}\,
S^{ik}_{}-\delta^{il}_{}\,S^{jk}_{}-\delta^{jk}_{}\,S^{il}_{})\ .
\ee 
The Hamiltonian-like boosts  are
\be 
M_{}^{-i}~=~x_{}^-p_{}^i-\frac{1}{2}\{x^i,P_{}^-\}+\frac{1}{p^+}
(T_{}^i-p_{}^jS_{}^{ij})\ ,
\ee
where the $T^i_{}$   transform as $SO(d-2)$ vectors
\be
[\,S^{ij}_{}\,,\,T^k_{}\,]~=~i\delta^{ik}_{}T^j_{}-i\delta^{jk}_{}T^l_{}\
,
\ee
and satisfy
\be[\,T^i_{}\,,\,T^j_{}\,]~=~im_{}^2S^{ij}_{}\ .\ee
When $m\ne 0$,  $T^i/m$ are the generators of
$SO(d-1)/SO(d-2)$, which together with  $S^{ij}$, complete  the  
massive little group $SO(d-1)$. 

When $m=0$, the $T^i$  commute with one another, acting  as light-cone
translations, and the algebra can be satisfied in two ways: 
\begin{itemize}
  \item $T^i=0$. This corresponds to the familiar massless representations
which describe particles with a finite number of degrees of freedom,
realized on states that  satisfy
\be
T^i\vert~p^+,p^i;\, (a_1,\dots,a_{r})\,>~=~0\ ,\ee
where $(a_1,\dots,a_r)$ are the Dynkin labels of  $SO(d-2)$
representations and $r$ is the rank of the little group. These label 
the different helicity states of the
massless particle. In four dimensions, the Pauli-Lubanski vector is
light-like.
\item $T^i\ne 0$. In this case, $T^i$ are the $c$-number components of a 
transverse vector. The states on which the Poincar\'e algebra 
is realized are
\be
T^i\vert~p^+,p^i;\, \xi^i,\, 
(a_1,\dots,a_{r})\,>~=~\xi^i~\vert~p^+,p^i;\,\xi^i,\,
(a_1,\dots,a_{r})\,>\ ,
\ee
which have additional labels, in the form of a little group vector
$\xi^i$. There is an important difference from the previous case, since $(a_1,\dots,a_{r})$ now labels  the $SO(d-3)$ subgroup of the 
transverse little group $SO(d-2)$. In four dimensions, there is no such group and the states are simply labelled by an additional space-like vector of constant magnitude.  
These  span two distinct representations,  called ``continuous 
spin representations" by Wigner in his original work {\cite{WIGNER}}. They are characterized by a space-like Pauli-Lubanski vector, and describe a massless state with an infinite number of integer-spaced helicities. 
\end{itemize}
In the following, we construct these representations in higher dimensions, as well as their supersymmetric generalizations.  

\subsection{Continuous Spin Representations}
These representations are characterized by the fact that the light-cone 
translations $T^i$ do not vanish,
even though $m=0$. It follows that a finite boost creates an infinite number of
integer-spaced helicities. To see this, consider a ket that represents a
state with light-like momentum aligned along the $z$ direction. In 
(3+1)-dimensions,  a rotation in the transverse plane yields
\be
e^{i\phi M^{12}}\vert~p^+,\,p^i=0;\, \Xi, \alpha>~=~
\vert~p^+,p^i=0;\, \Xi, \alpha+\phi>\ ,
\ee
where $\Xi$ is the length of the transverse light-cone translation vector 
with components 

$$ \xi^1~=~ \Xi \cos \alpha \ , \qquad \xi^2~=~ \Xi  \sin \alpha \ , $$
which enables us to set  

$$S^{12} = -i\frac{\partial}{\partial \alpha}\ .$$
Since $[ S^{12},\xi^i]\ne 0$ for $i=1,2$,  $S^{12}$ (helicity) 
is no longer a good quantum number; instead it acts on periodic or anti-periodic functions of the angle $\alpha$
$$
\vert\,\alpha\,> ~=~ \sum_{\lambda}\, e^{2\pi i\lambda \alpha}\,\vert\,f_\lambda\,> \ ,
$$
 where $\lambda$ are either integer-spaced integers or 
half-odd integers. Another way to see the infinite number of helicities is to apply the infinitesimal boost  $M^{-i}$ on any state

\be
M_{}^{-i}\vert~p^+,\,p^i=0; \Xi , \alpha>~=~
\frac{T^i}{p^+}\,\vert~p^+,\,p^i=0; \Xi , \alpha>\ ,
\ee
which produces a state which has picked up plus or minus one unit of
helicity. In a finite boost, this action is repeated an infinite amount,
resulting in states with possible helicities ranging from minus to plus
infinity in integer steps. There are (in four dimensions only) two types of representations,
those with all integer (single-valued)and those with all half-odd 
integer (double-valued) helicities. 
The square of the Pauli-Lubanski vector (Casimir) is the squared  length of the translation  vector
\be
W^\mu~W_\mu={\vec T}\cdot{\vec T} ~=~ \Xi ^2 \ .
\ee
As we have stated in the introduction, the CSRs have no obvious physical applications, explicitly in local field theories, but the appearance of an infinite number of states may indicate a connection with  non-local theories of extended
objects. This motivates their study in more general contexts. We explicitly construct these representations in higher dimensions, then proceed to show that they sustain supersymmetry and build continuous spin representations of supersymmetry. Finally we show that   if the light-cone translations 
are nilpotent, the CSRs become finite-dimensional, but at the expense of negative or zero-norm state, and the supersymmetric generalizations yield central charges for these massless representations.
\subsubsection{Higher Dimensions}
The front-form  of the Poincar\' e generators  shows that  CSR's in any dimensions, correspond to representations of the Galilean little group 

\be
[\,T^i\,,\,T^j\,]~=~0\ ,\qquad [\,S^{ij}_{}\,,\,T^k_{}\,]~=~i\delta^{ik}_{}T^j_{}-i\delta^{jk}_{}T^l_{}\ ,\ee
where the $T^i$ {\it do not vanish} and the $S^{ij}$ generate the light-cone little group $SO(d-2)$. Consider  $(4+1)$-dimensions, where the light-cone little group
is $SO(3)$. In the frame
$p^-=\vec p=0$, the Pauli-Lubanski two-form is

$$ W_{ij} ~=~ \frac{1}{\sqrt{2}}~\epsilon_{ijk} T^k \ ,$$
so that  one Casimir is 

\be W^{}_{ij}\,W^{}_{ij} ~=~ \Xi ^2 \ , \ee
as in the 4-dimensional case. The other Casimir is the Pauli-Lubanski zero-form

$$
W~=~\frac{1}{\sqrt{2}}~\epsilon_{ijk} 
T^i S^{jk} \ ,$$
which is  the projection of the generator 
$S^{jk}$ along $T^i$. The vector $T^i$ then acts as a ``quantization axis'', along which $W$ assumes the values 
\be 
\frac{W}{\sqrt{2}\,\Xi }~=~ 0, \, \pm \frac{1}{2} , \, \pm 1 , \, 
\pm \frac{3}{2}, \cdots \ .  
\ee
It follows that there are two types of representations corresponding to 
integer and half-odd integer values of $W/\sqrt{2}\,\Xi $. For each value 
of $W/\sqrt{2}\,\Xi $, 
there corresponds one infinite dimensional representation. Unlike $(3+1)$ 
dimensions, there are infinite numbers of CSRs in higher dimensions. The states are no longer characterized by the light-cone little group but by its subgroup orthogonal to $T^i$, which we call the short little group, $SO(2)$ in this case. Each CSR is
labeled by a  value of $SO(2)$ (integer of half-odd integer) as well as by the length of 
$ T^i$.

It is straightforward to find  the CSR representations in terms of $|\,j\,,\, m\,>$, the eigenstates of $SO(3)$, the full little group. Its states are  required to be 
eigenstates of $\vec T$, 
the $SO(2)$ rotations about it, and of the Casimir $\epsilon_{ijk} T^i S^{jk}$. 
As the action of $T^i$ on each $\vert\,j\,,\,m\,>$ yields a linear combination of $j=j,j\pm1$ states, its eigenstates are infinite linear combinations of eigenstates of the full little group
$SO(3)$. To see this in detail, take the light-cone vector in the $z$ direction, so that 
$\vec{T}$ is like the tensor operator $T^1_0 ~ \equiv ~ Y^1_0$, 

$$
T^1_0|\,j\,,\,m\,> ~= ~ \alpha^{(jm)}_+ |\,j+1\,,\,m\,> + \alpha^{(jm)}_0|\,j\,,\,m\,> 
+ \alpha^{(jm)}_-|\,j-1\,,\,m\,> \ ,
$$
where the $\alpha$'s are proportional to Clebsch-Gordan(C-G) 
coefficients

\begin{eqnarray} \nonumber 
\alpha^{(jm)}_0 & = &\Xi \, \sqrt{(j+m)(j-m+1)} \ , 
\quad -j<m<j \\ \nonumber 
\alpha^{(jm)}_\pm & = & \Xi \,\sqrt{(j^\prime +m)
(j^\prime -m +1)} \ ; \quad j^\prime = j\pm 1, 
\quad -j^\prime <m<j^\prime \ .
\end{eqnarray}
 The state of the form

$$|\,F >~=~\sum_{j=|M|}^\infty\,f^{(M)}_j\,|\,j\,,\,M\,>\ ,$$
is an eigenstate of the Casimir 

$$\frac{1}{2}\,\epsilon_{ijk} T^i S^{jk}\,|\,F\,>~=~\Xi\,M\,|\,F\,>\ ,~~~~~M=0,\pm \frac{1}{2},\pm 1,\dots\ ,$$ 
and of $T^3$, as long as the coefficients satisfy the recursion relations

$$\alpha_-^{(|M|+p,M)}\,f^{(M)}_{|M|+p}+\alpha_+^{(|M|+p-2,M)}\,f^{(M)}_{|M|+p-2}
+(\alpha_0^{(|M|+p-1,M)}-\Xi)\,f^{(M)}_{|M|+p-1}
~=~0\ ,$$
with $p=2,3,\dots$ and 

$$\alpha_-^{(|M|+1,M)}\,f^{(M)}_{|M|+1}
+(\alpha_0^{(|M|,M)}-\Xi)\,f^{(M)}_{|M|}
~=~0\ ,$$
since $f^{(M)}_{|M|-1}=0$.
The other states of the CSR are determined by acting the $SO(3)$ raising and lowering operators on $|\,F >$. Unlike the usual case, their action does not terminate, producing an infinite number of $SO(3)$ representations. This is similar to Wigner's CSR which contains an infinite number of integer spaced helicities. The difference in higher dimensions is 
that there is an infinite number of CSRs, each labelled by $M$.

\vskip .5cm

In $(5+1)$-dimension the light-cone little group is $SO(4)$, and there are two Pauli-Lubanski forms  
$$ \barr {rlcl}
{\rm 1-form :} & W_i & = & \frac{1}{\sqrt{2}}~\epsilon_{ijkl} 
T^j S^{kl} \ , \\ \\
{\rm 3-form :} & W_{ijk} & = & \frac{1}{\sqrt{6}}~\epsilon_{ijkl} T^l \ . 
\earr $$
The square of the highest form is $\Xi ^2$, a general feature in any dimensions. The square of one-form is the other Casimir

$$ W^{}_{i}\,W^{}_{i} ~=~ - \Xi ^2\, S^2_{\perp} \ , $$ 
where $S^{ij}_\perp$ are the generators of the $SO(3)$ subgroup of $SO(4)$
perpendicular to $T^i$,  and
$S^2_\perp \equiv S^{kl}_{\perp} S^{lk}_{\perp}~(k,l=1,2,3)$ is its
Casimir operator.  Thus

\be
W^{}_{i}\,W^{}_{i} ~=~  \Xi ^2\,j(j+1)\ ,\qquad j~=~ 0\ , \frac{1}{2}\ , 1\ ,  \frac{3}{2}\ ,  \cdots \ . \ee 
There are an infinite number 
of distinct CSR, one for each value of $j$. The states of the CSR can be labeled by finite
rotations of $SO(3)$ and vector $\vec{T}$ characterized by length, $\Xi $,
and three angles,
$$
 \vert~ p^+\ , p^i=0\ ; \Xi \ , \Omega_3\ ; j,m\,>  \ . 
$$
where $m$ ranges from $-j$ to $+j$ in integer steps.
\vskip .5cm

In $(6+1)$-dimension the little group is $SO(5)$ and the Pauli-Lubanski
forms are,
$$ \barr{rlcl}
{\rm 0-form :} & W & = & \frac{1}{2\sqrt{2}}~\epsilon_{ijklm} 
T^i S^{jk} S^{lm} \ , \\ \\
{\rm 2-form :} & W_{ij} & = & \frac{1}{2}~\epsilon_{ijklm} 
T^k S^{lm} \ , \\ \\
{\rm 4-form :} & W_{ijkl} & = & \frac{1}{2\sqrt{6}}~\epsilon_{ijklm} 
T^m \ .
\earr $$
The two Casimirs of the short little group $SO(4) \approx SU(2) \times SU(2)$, are given by 

\be
W^2_{(2-{\rm form}^{})} ~=~- \Xi ^2\, S^2_{\perp} \ , \ee
where here $S^{ij}_\perp$ are the generators of $SO(4)$ subgroup of
$SO(5)$ which leave $\vec{T}$ invariant, and the zero-form which gives the projection of the vector 
$\epsilon_{ijklm}S^{jk}S^{lm}$ along $\vec T$. The CSR states, characterized by 
four angles $\Omega_4$, and the $(j_1,j_2)$ representations of $SU(2)\times SU(2)$, 
 are of the form
$$ 
  \vert~ p^+\ , p^i\ ; \Xi \ , \Omega_4\ ; j_1\ ,m_1\ ;j_2\ ,m_2\,> 
$$
where $-j_{1(2)}\le m_{1(2)} \le j_{1(2)}$. There is an infinite number of CSRs, 
each labeled by $j_1$ and $j_2$.

The pattern is now clear: the CSRs are labeled by the length of a vector and by the irreps of the short little group $SO(d-3)$ which leaves it invariant. Its Casimirs are directly related to the squares of the Pauli-Lubanski $n$-forms and the  zero-form. There are infinitely many such representations,  fermionic and  bosonic. 

The case of eleven space-time dimensions is particularly interesting because of M-theory and supergravity. Its CSR's are labelled by $SO(8)$, with its magic triality property which acts on 
 its vector representation, ${\bf 8}_v$, and two inequivalent spinor 
representations, ${\bf 8}_s$ and ${\bf 8}_s'$.

We conclude this section by writing the general expressions for the Casimir operators of the Poincar\' e algebra in terms of those of the short little group.   The Pauli-Lubanski $n$-form for 
the light-cone little group $SO(d-2)$ is given by Eq.(\ref{forms}),
$$
W_{i_1 \cdots i_n} ~=~ 
\frac{ \epsilon_{i_1 \cdots i_n
i_{n+1} \cdots i_{d-2}} T^{i_{d-2}} S^{i_{n+1} i_{n+2}} 
\cdots S^{i_{d-4} i_{d-3}}}{\sqrt{n!~ 2_{}^{\frac{d-n-3}{2}}
\left(\frac{d-n-3}{2}\right)!}} \ .
$$
The Casimir operators are simplest to calculate in a frame where only $T^{i_{d-2}} \ne 0$, and the square of the highest form is the squared length of the vector $\vec T$.  The Casimir operators for
$n\ge 1$ are given by,
\begin{itemize}
\item for little group $SO(6)$ and $SO(7)$ :
$$ 
W^2_{i_1 \cdots i_n} ~=~ 
\Xi ^2 \left\{ (S^2_\perp)^{(d-n-3)/2} -2 (S^{(d-n-3)}_\perp)\right\}
\ ; \ 5 \ge d-n-2 > 1
\ .
$$
\item For little group $SO(8)$ and $SO(9)$ :
$$ 
W^2_{i_1 \cdots i_n} ~ =~
\Xi ^2 \left\{ \begin{array}{ccl}
(S^2_\perp)^{(d-n-3)/2} -2 (S^{(d-n-3)}_\perp)
& ; & 1<d-n-2 <7 \\ \\
-(S^2_{\perp})^3 + 2 S^2_\perp S^4_{\perp}
- 2 S^6_\perp & ; & d-n=9
\end{array}\right.
$$

\item In general for the little group $SO(d)$ :
$$
W^2_{i_1\cdots i_n} ~ =~ \Xi ^2 
\sum\limits^{k}_{p=0} \left(A_{2p} S^{2p}_\perp S^{2(k-p)}_\perp 
+ B_{2p} (S^2_\perp)^k + C_{2p} (S^2_\perp)^p S^{2(k-p)}_\perp \right) \ , 
$$
where $k= \frac{1}{2}(d-n-3)$, and $A_{2p}$, $B_{2p}$ and $C_{2p}$'s
are numerical constants. In the above, $S^{ij}_\perp$ are the generators of $SO(d-3)$ subgroup of $SO(d-2)$ 
perpendicular to $T^{i_{d-2}}$ and 
$$ S^{m}_{\perp} \equiv S^{i_1 i_2}_\perp S^{i_2 i_3}_\perp 
\cdots S^{i_{m-1}i_m}_\perp S^{i_m i_1}_\perp \ . 
$$
\end{itemize}
In odd dimensions, the extra Casimir operator is provided by the  Pauli-Lubanski zero-form
$$
W_{(0)} ~=~ \frac{1}{\sqrt{2^{\frac{d-3}{2}}  \;
\left(\frac{d-3}{2}\right)!}}   
~ \epsilon_{ijk \cdots mn}T^i S_\perp^{ij} \cdots S_\perp^{mn} \ .
$$
In higher dimensions, the CSR states are  
labeled by  $\Xi $, and the solid angle in $d-3$ dimensions, $\Omega_{d-3}$ which give the length and direction of $\vec T$, respectively, as well as by 
$(a_1,\cdots , a_r)$, the Dynkin labels of $SO(d-3)$

$$ \vert ~p^+\ ,\, p^i\ ;\,\Xi \ ,\Omega_{d-3}\ ;\,(a_1,\cdots , a_r)\, >\ .
$$
These differ from the usual massless representations in that they are 
characterized by a space-like vector, and contain an infinite number of states. 

\section{Super-Poincar\'e Algebra}
\subsection{Super-Charges in Light-Cone Form}
The continuous spin representations can be generalized to include
supersymmetry. Indeed, Wigner originally found two CSR's, one  single-valued with
integer helicities, the other  double-valued with half-odd integer
helicities. As we shall see, they are related by supersymmetry.

The super-Poincar\' e algebra includes the  spinor generators which satisfy 

\bea \label{sp1}
\left[Q_A \ , \ p^\mu \right] & = & 0 \ ,  \\ \label {sp2}
\left[M^{\mu\nu} \ , \ Q_A \right] & = & -\frac{1}{2} \left(
\Sigma^{\mu\nu} Q \right)_A \ , \\ \label{sp3}
\{ Q_A \ , \ Q^\dag_B \} & = & (\gamma^\mu p_\mu 
\gamma^0)_{AB}  \ ,
\eea
where $A,B$ are spinor indices. In the above, 

$$ \Sigma^{\mu \nu} = -\frac{i}{2} [ \gamma^\mu \ , \ \gamma^\nu ]
\ , $$
where the $\gamma$ matrices satisfy the anticommutation relation,

$$ \left\{ \gamma^\mu \ , \ \gamma^\nu \right\} \ = \ 2 \eta^{\mu \nu} 
\ ; \qquad {\mu , \nu = 0, \cdots ,d}  \ .$$
We first consider these relations in four space-time dimensions. 
The supercharges do not commute with the Pauli-Lubanski vector, as

$$[W_\mu, Q_A]~=~\frac{i}{2}p^\nu(\Sigma_{\mu\nu}\gamma_5 Q)_A\ ,$$
using

$$
\frac{1}{2}\epsilon_{\mu\nu\rho\sigma}\Sigma^{\rho\sigma}=
-i\Sigma_{\mu\nu}\gamma_5\ ,
$$
and $\gamma^5 = -i \gamma^0 \gamma^1 \gamma^2 \gamma^3$ is the chirality matrix. 
The supercharges can be realized linearly using Grassmann variables 
and their derivatives as \cite{BBB},

\be \label{sp4}
Q_A = {\partial}_A + \frac{1}{2} (\gamma^\mu  p_\mu
\gamma^0)_{AB} \bar{\theta}^B \ , \nonumber
\ee
and its conjugate as

\be \label{sp5}
Q_C^\dag = \bar{\partial}_C + \frac{1}{2} (\gamma^\mu 
p_\mu \gamma^0)_{DC} {\theta}^D \ ,
\ee
where we used 

$$\gamma^0 \gamma^{\mu \dag} \gamma^0 = \gamma^\mu\ ,\nonumber $$
and $\partial_A = \frac{\partial}{\partial \theta^A}$,
$\bar{\partial}_A = \frac{\partial}{\partial \bar{\theta}^A}$, and
$\bar{\theta}_A$ is the complex conjugate of $\theta_A$ with
${A,\cdots , D}$ running over the spinor indices. $\theta$, $\bar{\theta}$, 
$\partial$ and
$\bar{\partial}$ are anticommuting Grassmann parameters.
 We use the light-cone projectors to split the supercharge into
two-component spinors,

$$ Q = Q_+ + Q_- \ , \qquad {\rm where} 
\quad Q_\pm \equiv {\cal P}_\pm Q \ ,$$
with ${\cal P}_\pm=\frac{1}{2}\gamma^\mp\gamma^\pm$. In the Weyl representation, from Eq.(\ref{sp3}-\ref{sp5}), 
we can now read off the various anticommutation relations,
namely,
\bea
\left\{ Q_{+a} \ , \ Q_{+b}^\dag \right\} & = & \sqrt{2}\; p^+ \; 
\delta_{ab} \ , \nonumber \\
\left\{ Q_{-a} \ , \ Q_{-b}^\dag \right\} & = &
\sqrt{2}\; p^-\; \delta_{ab} ~\equiv~ 
\sqrt{2}\; \frac{p \bar{p}}{p^+}\; \delta_{ab} \ , \nonumber \\
\left\{ Q_{-1} \ , \ Q_{+1}^\dag \right\} & = & -\sqrt{2}\;\bar{p} \ , \qquad
\left\{ Q_{+1} \ , \ Q_{-1}^\dag \right\} ~=~ -\sqrt{2}\;p \ , \nonumber \\
\left\{ Q_{+2} \ , \ Q_{-2}^\dag \right\} & = &\ \ \sqrt{2}\;\bar{p} \ , \qquad
\left\{ Q_{-2} \ , \ Q_{+2}^\dag \right\} ~=~ \ \ \ \, \sqrt{2}\;p \ ,
\nonumber 
\eea
where $p(\bar{p}) = \frac{1}{\sqrt{2}}(p^1 \pm i p^2)$ and $a,b=1,2$. All 
other 
anticommutators being zero. This shows that 
the supercharge
splits up into two disconnected sets of anticommutators, corresponding to
the left and right projections (${\cal P}_{L,R}=\frac{1}{2}(1\pm \gamma_{5})$)
\be {\cal Q}^{ R}_\pm \equiv  {\cal P}_R Q_\pm\ ,\qquad {\cal Q}^{\rm
L}_\pm \equiv  {\cal P}_L Q_\pm\ .\nonumber\ee
One could also think of these as representing $N=2$ supersymmetry. 

There is further reducibility, indicated by the fact that 
the supercharge anticommutes with the covariant derivative
\be\nonumber
{\cal D}_A = {\partial}_A - \frac{1}{2} (\gamma^\mu  p_\mu
\gamma^0)_{AB} \bar{\theta}^B \ , \label{cov}
\ee
and their hermitian conjugates, which give the following relations,
\be
{\cal Q}^L_{-}~=~-\frac{\bar p}{p^+}\;{\cal Q}^L_{+}\ ,\qquad {\cal
Q}^R_{-}~=~\frac{ p}{p^+}\;{\cal Q}^R_{+}\ ,\nonumber
\ee
together with their conjugates
\be
{\cal Q}^{L~\dagger}_{-}~=~-\frac{ p}{p^+}\;{\cal Q}^{L~\dagger}_{+}\
,\qquad {\cal Q}^{R~\dagger}_{-}~=~\frac{\bar p}{p^+}\;{\cal
Q}^{R~\dagger}_{+}\ ,\nonumber
\ee
without affecting the anticommutation relations. 
From now on we concentrate on the right-handed
projections of the algebra. On (right-handed) superfields, the
constraint becomes
\be
{\cal P}_R{\cal D}\;\Phi~=~0\ ,\nonumber
\ee
and after using mass-shell condition, 
we are left with only one supercharge, $\cal Q_+$ (henceforth we drop the
superscript $R$). It is expressed 
in terms of one Grassmann variable $\theta_3\equiv \theta$ and its
conjugate, as we can drop the $\theta_4$ variable altogether since its
derivative is now expressed in terms of the derivative with respect to
$\theta_3$. Hence 
\be
{\cal Q}_+~=~\frac{\partial}{\partial
\theta}+\frac{1}{\sqrt{2}}p^+\overline\theta\ ,\qquad
{\cal Q}^\dagger_+~=~\frac{\partial}{\partial
\overline\theta}+\frac{1}{\sqrt{2}}p^+\theta\ .
\ee
Since 
\be
[M^{+-},{\cal Q}_{\pm}] ~=~\pm\frac{i}{2}\,{\cal Q}_\pm\ ,
\ee
we must extend the Bacry-Chang representation of $M^{+-}$ to 
\be
M^{+-}~=~-x_{}^-p_{}^+-\frac{i}{2}\,\left(\theta\frac{\partial}
{\partial\theta}+\overline\theta\frac{\partial}{\partial\overline\theta}
\right)\ .
\ee
In order to satisfy the other commutation relations
\be
[M^{+i},M^{-j}]~=~i\delta^{ij}M^{+-}-iM^{ij}\ ,\qquad [M^{12},{\cal
Q}_{\pm}] ~=~\mp\frac{1}{2}\,{\cal Q}_\pm\ ,
\ee
the generators $M^{12}$ and $M^{-i}$ now include 
the $\theta$-dependent terms, 
\be
M^{12}~=~x^1p^2-x^2p^1+\widehat S^{12}_{} \ ,
\ee
as well as 
\be 
M_{}^{-i}=x_{}^-p_{}^i-\frac{1}{2}\{x^i,P_{}^-\}+\frac{1}{p^+}
(T_{}^i-p_{}^j\widehat S_{}^{ij})+i\frac{p^i}{2p^+}
\left(\theta\frac{\partial}
{\partial\theta}+\overline\theta\frac{\partial}{\partial\overline\theta}
\right)\ ,
\ee
where now 
\be
\widehat S^{12}~=~S^{12}_{} 
+\frac{1}{2}\,\left(\theta\frac{\partial}{\partial\theta}-\overline\theta
\frac{\partial}{\partial\overline\theta}\right)\ .
\ee
The generator $M^{+i}$ remains unchanged. 
The other relevant commutators in light-cone form are
\be [M^{\pm i},{\cal Q}_\pm ]~=~ 0 \ .\ee
In the frame where $p^i=0$, we have  
\be{\cal Q}_-=0\ ,\qquad M^{-i}=T^i/p^+\ ,\ee
so that the commutators
\be
[M^{-1},{\cal Q}_{+}] ~=~-\frac{i}{\sqrt{2}}\,{\cal Q}_-\qquad 
[M^{-2},{\cal Q}_{+}] ~=~-\frac{1}{\sqrt{2}}\,{\cal Q}_-\ ,
\ee
\be 
[M^{+1},{\cal Q}_{-}] ~=~-\frac{i}{\sqrt{2}}\,{\cal Q}_+\ ,\qquad 
[M^{+2},{\cal Q}_{-}] ~=~\frac{1}{\sqrt{2}}\,{\cal Q}_+\ ,\ee
imply that
\be[T^i,{\cal Q}_+]~=~0\ .\ee
Since the vector that characterizes the CSRs commutes with the supercharge, we can  
implement supersymmetry on the continuous spin representations without
having to change the supercharges. 

Thus in $(3+1)$ dimension, we obtain the
{\it unique} representation of supersymmetry which contains all  integer and
half-odd integer  helicities. This of course does not alleviate  the  problems 
associated with the CSRs.
\subsection{Higher Dimensions}
In higher dimensions, Eqs.(\ref{sp1}-\ref{sp5}) still hold, but 
the number and nature of independent Grassmann
parameters, i.e., whether these are Dirac, Weyl or Majorana type, depend
on the number of space-time dimensions\cite{SOHNIUS}. In $d$-dimensions,
there are $2^{d_{}/2}(2^{{(d-1)}_{}/2})$ complex spinor
components for $d$ even(odd).
Using the  anticommutativity between
supercharge and covariant derivative, these numbers can be further
reduced by a factor of $2$. Let $a,b$ run over the independent spinor
indices, i.e., $a,b=1,2,\cdots , 2^{d/2 -m}(2^{(d-1)/2-m})$, for $d$
even(odd) respectively, where $m$ is the number of independent constraints. 
Using all reducibility
conditions, we can always express the super-Poincar\'e algebra in the
light-cone form,
\bea \label{m(ij)}
M^{ij} & = & x^i p^j - x^j p^i + \widehat S^{ij} \ , \\ \label{m(+-)}
M^{+-} & = & -x^- p^+ + S^{+-} \ , \\ \label{m(-i)}
M^{-i} & = & x^- p^i -\frac{1}{2} \{ x^i , P^- \} + \frac{1}{p^+}(T^i -
p^j \widehat S^{ij} ) - \frac{p^i}{p^+} S^{+-} \ , 
\eea
where
\bea \label{s(ij)}
\widehat S^{ij} & = & S^{ij} + \left( \frac{1}{2} \theta^a
(\gamma^{ij})_{ab} \partial^b + c.c. \right) \ , \\ \label{s(+-)}
S^{+-} & = & \left( \frac{1}{2}\theta^a (\gamma^{+-})_{ab} \partial^b 
+ c.c. \right) \ , 
\eea
and $\gamma^{ij}$ and $\gamma^{+-}$ are the reduced Dirac submatrices
consistent with the constraints. $M^{+i}$ remain the same as
before. The indices $i,j$ run over transverse space.
The supercharges can be irreducibly expressed as,
\be \label{higher-dim}
\cq^a_+ ~=~ \frac{\partial}{\partial \theta^a} + \frac{1}{\sqrt{2}} p^+
\bar{\theta}^a \ , \qquad
\cq^{\dag a}_+ ~=~ \frac{\partial}{\partial \bar{\theta}^a} +
\frac{1}{\sqrt{2}} p^+ \theta^a \ , 
\ee
and $\cq_-^a$ can be expressed in terms of $\cq_+^a$. To illustrate this construction, we now show how supersymmetric CSRs arise in five and eleven dimensions:
\subsubsection{(4+1)-dimensions}
In five dimensions, the spinors have four complex components which can be 
reduced to two using the  covariant derivative constraint. 
Using the light-cone projectors, ${\cal P}_\pm = -\frac{1}{2}\gamma^\mp
\gamma^\pm$ in the representation,
\be \gamma^0 = i \sigma^1 \otimes I \ , \ \gamma^i = \sigma^3 \otimes
\sigma^i \ , \ {\rm i = 1,2,3} \ ; \ {\rm and} \ \gamma^4 = \sigma^2
\otimes I \ ,
\ee
and Eq.(\ref{sp3}), we find
\bea
\left\{ \cq_{+a} \ , \ \cq_{+b}^\dag \right\} & = & \sqrt{2} \, p^+
\delta_{ab} \ , \\
\left\{ \cq_{-a} \ , \ \cq_{-b}^\dag \right\} & = & \sqrt{2} \, p^-
\delta_{ab} \
= \ \ \frac{\vec{p} \cdot \vec{p}}{\sqrt{2} \, p^+} \delta_{ab} \ , \\
\left\{ \cq_{+a} \ , \ \cq_{-b}^\dag \right\} & = & i
\left(\vec{\sigma}
\cdot \vec{p} \right)_{ab} \ , \\
\left\{ \cq_{-a} \ , \ \cq_{+b}^\dag \right\} & = & -i
\left(\vec{\sigma}
\cdot \vec{p} \right)_{ab} \ ,
\eea
\noi where $a,b=1,2$ and also we used the on-shell condition
$p^- = {p^i p^i}/{2p^+}$.
From Eq.(\ref{cov}) and ${\cal D}\Phi =0$, we find,

$$ 
\frac{\partial}{\partial \theta^{2+a}} ~=~ - i \left( \frac{\vec{\sigma}
\cdot \vec{p}}{\sqrt{2} \, p^+}\right)_{ab} 
\frac{\partial}{\partial \theta^b} \  , 
$$
and similarly for their complex conjugates, allowing us to set
$(\theta^3,\theta^4)=0$. The supercharges, Eq.(\ref{sp4}-\ref{sp5}), 
reduce to the irreducible form as in Eq.(\ref{higher-dim}) with $a,b=1,2$, and 

\be Q_{-a} = -i \left( \frac{\vec{\sigma} \cdot \vec{p}}{ \sqrt{2}
\, p^+} Q_+ \right)_a \quad {\rm and \quad hence} \quad Q_{-a}^\dag =i
\left( Q_+^\dag \frac{\vec{\sigma} \cdot \vec{p}}{ \sqrt{2}   
\, p^+} \right)_a \ .
\ee
The Lorentz generators remain the same as in 
Eqs.(\ref{m(ij)}-\ref{m(-i)}) with the following irreducible representations 
of $S^{ij}$ and $S^{+-}$(in Eq.(\ref{s(ij)}-\ref{s(+-)})),
\bea
S^{+-} & = & -\frac{i}{2}\left( \theta \frac{\partial}{\partial \theta} 
+ \bar{\theta} \frac{\partial}{\partial \bar{\theta}} \right) \ , \\ 
\widehat S^{ij} & = & S^{ij} + \frac{1}{2} 
\epsilon^{ijk}\left(\theta
\sigma^k \frac{\partial}{\partial \theta} + c.c. \right) \ ,
\eea
where $\theta=(\theta^1,\theta^2)$ and
$\frac{\partial}{\partial \theta}=\left(\frac{\partial}{\partial
\theta^1},\frac{\partial}{\partial \theta^2}\right)$. The supercharges 
transform as  $SO(3)$ spinors,
\be
[M^{ij} \ , \ \cq_{+a}] ~=~ -\frac{1}{2} \epsilon^{ijk}(\sigma^k
\cq_+)_a \ .
\ee
These charges can be made to act on the CSRs, 
for which the relevant group is that of the short little group $SO(2)$ 
which leaves the light-cone translation vector invariant. Aligning 
$\vec T$ along the z-axis,  the supercharges split into 
two components, $\cq_{+1}$ and $\cq^\dag_{+2}$, that lower the value 
of $M^{12}$ by half a unit, and two components, $\cq_{+2}$ and 
$\cq^\dag_{+1}$ that raise it. 

In terms of $M$, the eigenvalue of $\widehat S^{12}$, the supermultiplet 
consists of 

\bea \nnum 
|\,M>_{\csr}&\sim &|\,M>_{\csr}\ ,\\\nnum
\cq_{+1}|\,M>_{\csr}, \  \cq^\dag_{+2}|\,M>_{\csr} & 
\sim & |\,M - \frac{1}{2}>_ {\csr} \ , \\ \nnum
\cq_{+1} \cq_{+2}^\dag|\,M>_{\csr} & \sim & |\,M -1>_{\csr} \ .
\eea
It contains two bosonic and two fermionic CSR, with the same structure as the 
ordinary $(\vec{T}=0)$ massless $N=2$ supermultiplet in four dimensions. The important difference 
is that it contains not only the ordinary states but their copies under the boosts proportional to $\vec T$. This yields as usual an infinite number of $SO(3)$ polarization states. The action of supersymmetry is the same as in the normal case, but  the CSR supermultiplets contain an infinite number of ordinary massless supermultiplets of ever-increasing spin. 
\subsubsection{(10+1)-dimensions} 
The case of eleven dimensions is particularly interesting because it is 
shrouded in mystery, as it contains not only local supergravity but 
also the elusive M-theory. What we know of M-theory is that its 
compactifications to lower dimensions yields supersymmetric theories 
and that its long distance limit is the supergravity theory. 

In eleven dimensions, the spinors have 32 complex components, which 
upon  using the Majorana condition and the  covariant derivative 
constraint, reduce to 16 real components. The supercharge splits as

$$ Q ~=~ \left(\matrix{Q^A \cr  Q^{16+A}}\right) ~\equiv ~\left(\matrix{Q^A_+ \cr  Q^A_-}\right) \ , \quad A=1,\cdots , 16 \ . $$
The Majorana condition implies,

$$ Q^A_+ ~=~ Q^{\dag A}_+ \ , \quad {\rm and} \quad Q^A_- ~=~ 
- Q^{\dag A}_- \ . $$
In the representation
\be
\gamma^0=i\sigma^1 \otimes I, \quad \gamma^i = \sigma^3 \otimes 
{\tilde{\gamma}}^i , \quad \gamma^{10} = \sigma^2 \otimes I \ ,
\ee
where $i=1,\cdots , 9$ and ${\tilde{\gamma}}^i$'s are $16\times 16$ and real, 
symmetric matrices which satisfy the following algebra
$$ \{ {\tilde{\gamma}}^i , {\tilde{\gamma}}^j \} ~=~ 2\; \delta^{ij} \ , $$
the super-Poincar\'e algebra becomes
\bea \nnum
\{ Q^A_+ , Q^B_+ \} & = & \sqrt{2} \, p^+ \delta^{AB} \ , \\ \nnum
\{ Q^A_- , Q^B_- \} & = & -\sqrt{2}\,p^- \delta^{AB}~=~ -\frac{\vec{p}\cdot
\vec{p}}{\sqrt{2} \, p^+} \delta^{AB} \ , \\ \nnum
\{ Q^A_+ , Q^B_- \} & = & -i \tgamma^{AB}_i p^i \ , \\ \nnum
[ M^{ij}, Q^A_{\pm} ] & = & \frac{i}{2} (\tgamma_{ij} Q_\pm)^A \ , \\ \nnum
[ M^{+-}, Q^A_{\pm} ] & = & \pm\frac{i}{2}Q_\pm^A \ , \\ \nnum
[ M^{\pm i}, Q^A_{\pm} ] & = & 0 \ , \qquad
[ M^{\pm i}, Q^A_{\mp} ] ~=~ \pm \frac{i}{\sqrt{2}} 
(\tgamma_{i} Q_\pm)^A \ .
\eea
Using ${\cal D} \Phi=0$ and Eq.({\ref{cov}), we find,

$$ 
\frac{\partial}{\partial \theta^{16+A}} ~ = ~ -i \; \left(\frac{\tgamma^i p^i}
{\sqrt{2}\, p^+} \frac{\partial}{\partial \theta}\right)_A \ , 
$$
allowing us to set $\theta^{16+A}=0$ for 
$A=1, \cdots , 16$ . The remaining real $Q^A_+$'s  can be expressed 
in terms of 16 real Grassmann parameters,

$$Q^A_+ ~=~ \frac{\partial}{\partial \theta^A} + 
\frac{1}{\sqrt{2}} p^+ \theta^A  \qquad {\rm and} \quad
Q^A_- ~=~ -i \left(\frac{\tgamma^i p^i}{\sqrt{2}\, p^+} Q_+\right)^A \ .
$$
Similarly, in this representation, Eqs.(\ref{s(ij)}-\ref{s(+-)}) reduce to 

\bea \nnum 
S^{+-} & = & -\frac{i}{2}\; \theta^A \frac{\partial}{\partial
\theta^A} \ , \\ \nnum
\widehat S^{ij} & = & S^{ij} - \frac{i}{2} \; \theta^A
\tgamma^{ij}_{AB} \frac{\partial}{\partial \theta^B} \ ,
\eea
which together with Eqs.(\ref{m(ij)}-\ref{m(-i)}) give the Lorentz generators, 
where $\tgamma^{ij}~=~ \tgamma^i \tgamma^j$ for $i\ne j$.

We can let these charges act onto CSRs, remembering that they are labeled by 
the short little group $SO(8)$ that leaves the light-cone translation vector  $\vec{T}$ 
invariant.
We decompose the supercharges into two 8-component supercharge, $\cq^a_+$ 
and $\cq^{\dot{a}}_+$, by the  $SO(8)$ chirality matrix,

$$ 
Q^A_+ ~=~ \left(\matrix{\cq^a_+ \cr  \cq^{\dot{a}}_+}\right) \ , 
\qquad a=1,\cdots , 8
$$
where

$$\cq^a_+ ~=~ \frac{1}{2}(1+\tgamma^{(9)}) \quad {\rm and} \quad 
\cq^{\dot{a}}_+ ~=~ \frac{1}{2}(1-\tgamma^{(9)}) \ , 
$$
and the chirality matrix of $SO(8)$ subgroup is

$$
\tgamma^{(9)} \equiv \tgamma^1 \tgamma^2 \tgamma^3 \tgamma^4 \tgamma^5 
\tgamma^6 \tgamma ^7 \tgamma^8 \ . 
$$
These two 8-component supercharges furnish two inequivalent spinor 
representations of $SO(8)$, ${\bf 8}_s$ and ${\bf 8}^\prime_s$ 
characterized by opposite $SO(8)$ chirality and $\cq^a_+ \sim {\bf 8}_s , \ 
\cq^{\dot{a}}_+ \sim {\bf 8}^\prime_s$.

The $(16\times 16)$ $\tgamma^i$ matrices can be written in terms of 
$8\times 8$ block form
$$
\tgamma^i ~=~ \left(\matrix{ 0 & \tgamma^i_{a\dot{a}} \cr 
\tgamma^i_{\dot{b}b} & 0}\right) \ , 
$$
where $\tgamma^i_{a\dot{a}}$ is the transpose of $\tgamma^i_{\dot{a}a}$. The 
Clifford algebra for $\tgamma^i$ is satisfied if
$$
\tgamma^i_{a\dot{a}} \tgamma^j_{\dot{a}b} + 
\tgamma^j_{a\dot{a}} \tgamma^i_{\dot{a}b} ~=~ 2 \delta^{ij}\delta_{ab} \ , 
\quad {\rm for} \quad i,j,a,b=1,\cdots , 8
$$
and similarly with undotted and dotted indices interchanged. We also define
$$\tgamma^{ij}_{ab} ~=~ \frac{1}{2}(\tgamma^i_{a\dot{a}} 
\tgamma^j_{\dot{a}b} - \tgamma^j_{a\dot{a}} 
\tgamma^i_{\dot{a}b}) \ , 
$$
and similarly for $\tgamma^{ij}_{\dot{a}\dot{b}}\;$ . The $(8\times 8)$ 
$\tgamma^i_{a\dot{a}}$ matrices couple the vector and spinor representations.
 In this chiral subspace, the supercharges, $\widehat S^{ij}$ and 
$S^{+-}$ take the following irreducible forms
\bea
\cq^a_+ & = & \frac{\partial}{\partial \theta^a} + \frac{1}{\sqrt{2}} 
p^+ \theta^a \ , \\
\cq^{\dot{a}}_+ & = & \frac{\partial}{\partial \theta^{\dot{a}}} + 
\frac{1}{\sqrt{2}} p^+ \theta^{\dot{a}} \ , \\ \nnum 
\cq^a_- & = & -\frac{ip^i}{\sqrt{2} \, p^+} 
\tgamma^i_{a\dot{a}} \cq^{\dot{a}}_+ \ , \\ \nnum 
\cq^{\dot{a}}_- & = & -\frac{ip^i}{\sqrt{2} \, p^+} 
\tgamma^i_{\dot{a}a} \cq^a_+ \ ,  \nnum 
 \eea
as well as

\be
S^{+-} ~=~ -\frac{i}{2}\; \theta^a \frac{\partial}{\partial
\theta^a} \ , \qquad
\widehat S^{ij} ~=~ S^{ij} - \frac{i}{2} \; \theta^a
\tgamma^{ij}_{ab} \frac{\partial}{\partial \theta^b} \ .\ee
We have two supersymmetries, each transforming as a different $SO(8)$
spinor, and the basic supermultiplet is of the form 
$({\bf 8}_v + {\bf 8}_s)_{\csr} \times ({\bf 8}_v + {\bf 8}^\prime_s)_{\csr}$,  
 with 128 bosonic and 128 fermionic states, 
\bea \nnum
& &({\bf 8}_v + {\bf 8}_s)_{\csr} \times ({\bf 8}_v + {\bf 8}^\prime_s)_{\csr} 
\\ & = & 
({\bf 1}+{\bf 28} +{\bf 35} +{\bf 8} + {\bf 56})_{v,\csr} + 
({\bf 8}^\prime + {\bf 56}^\prime +{\bf 8}+{\bf 56})_{s,\csr}\ .
\eea
This supermultiplet (without CSR) is of course that of the massless states 
of IIA string theory, obtained by dimensional reduction from eleven-dimensional 
supergravity.  

As we found in the five-dimensional case, there is a  one-to-one correspondence between 
the labels of the ordinary massless representation for N=2 supersymmetry in ten dimensions 
and those of the massless CSR for N=1 supersymmetry in eleven dimensions. However the CSR supermultiplet contains the states of $N=1$ supergravity as well as an infinite number of massless supermultiplets obtained by boosting along $\vec T$.

\subsection{Dimensional Reduction}
The characteristic feature of  the CSRs is the transverse space vector
$T^i$. A non-zero value  implies an infinite number of polarization states while a zero value requires a finite number of polarizations. In covariant terms, this vector is written in terms of the  Pauli-Lubanski $(d-3)-$form, its magnitude unchanged by Poincar\'e transformations. 

This suggests several ways in which CSRs might play a role in Physics. One is to enlarge the invariance group to include transformations capable of changing the length. The other is to view the CSR in the context of dimensional reduction by limiting transformations to rotations perpendicular to $\vec T$.

The former approach requires a study of the representations of the
larger groups. Of prime interest are the conformal group (since we are
dealing with massless representations) and perhaps the de Sitter groups. In particular we need to identify how CSRs and normal massless representations reside in unitary irreducible representations of these groups. 

In physical terms we may  view the length as an order parameter
(as in the Higgs mechanism with a vector representation). For this we need to give it a dynamical meaning and couple it to an external field; it is amusing to note that in eleven dimensions it naturally couples to an eight-form, which is dual to a $3$-form. 

As we let the length tend to zero, we cannot  a priori determine 
which normal representation will be singled out in the transition. This is reminiscent 
of Dashen's theorem in chiral theories where an external additional tag is required; otherwise it points to a first-order phase transition. 
 
In the context of dimensional reduction, the translation vector $\vec T$ naturally singles out a subspace perpendicular to its direction. It follows that group operations restricted to that subspace span normal representations of the Poincar\' e and super-Poincar\' e algebras. Thus it 
might be possible to start with a problematic theory and dimensionally reduce it to a well-defined one.

\subsection{Nilpotent Light-Cone Translations}
We have seen that  CSRs  necessarily contain massless states of unbounded spins related by finite boosts, raising many objections for their use in the description of point-like objects. 
 On the other hand, the extension of Poincar\'e invariance to supersymmetry introduces nilpotent Grassmann variables, which  suggests the construction of  
the light-cone translations out of these  
Grassmann parameters. The translations would be nilpotent, and therefore  
generate a finite number of helicities with finite boosts. 

In (3+1)-dimensions, we need two complex Grassmann variables, 
$\theta_1\ ,\theta_2$, in order to build an $SO(2)$ vector. We set 

\be
T^1+iT^2~=~\sqrt{2}\,\overline z\,p^+\theta_1\;\theta_2\ ,\qquad
T^1-iT^2~=~\sqrt{2}\, z\,p^+\overline\theta_1\;\overline\theta_2\ ,
\ee
where, $z$ is a complex
variable, and we have added the appropriate power of $p^+$ to ensure
proper commutation with $M^{+-}$.  $S^{12}$ now becomes 

\be 
\widehat S^{12} \equiv S^{12} + \frac{1}{2}
(\theta^a\partial_a-\bar{\theta}^a\bar{\partial}_a) \ , 
\ee
and the Poincar\' e Casimir operator is now a nilpotent Grassmann number,

\be
W^2~=~W^\mu\,W_\mu~=~-2\,|z|^2(p^+)^2\,\theta_1\theta_2\overline\theta_1
\overline\theta_2\ .
\ee
With two Grassmann variables, we can construct  two supercharges,
corresponding to $N=2$ supersymmetry. The kinematic supersymmetries are  
unaltered 
\be
{\cal Q}_+^a~=~\frac{\partial}{\partial\theta_a}+\frac{1}{\sqrt{2}}p^+
\overline\theta_a\ ;\qquad
{\cal Q}_+^{a\,\dagger}~=~\frac{\partial}{\partial\overline\theta_a}
+\frac{1}{\sqrt{2}}p^+\theta_a\ ,
\ee
where $a=1,2$. However the light-cone translations no longer commute with these supercharges. To restore the super-Poincar\' e algebra, the 
dynamic supersymmetries must be altered to the new form
\be
{\cal Q}^a_{-}~=~\frac{ p}{p^+}\;{\cal Q}^a_{+}-i\overline
z\epsilon^{ab}\theta_b\ ,\qquad
{\cal Q}^{a~\dagger}_{-}~=~\frac{\bar p}{p^+}\;{\cal
Q}^{a~\dagger}_{+}+iz\epsilon^{ab}\overline\theta_b\ ,
\ee
which ensures  the commutation relations of ${\cal Q}_+^a$ with the boosts
$M^{-i}$. The resulting supersymmetry algebra 
\be
\{{\cal Q}_+^a\,,\, {\cal Q}_-^b\}~=~i\epsilon^{ab}\,\overline z\ ,\qquad
 \{{\cal Q}_+^{a\,\dagger}\,,\, {\cal 
Q}_-^{b\,\dagger}\}~=~-i\epsilon^{ab}\, z\ ,
\ee
acquires central charges, even though we are  in a massless representation.

This leads to  zero or negative norm states. The positive norm  states are
of the form  $\vert\cdots>$,  and ${\cal Q}^{a\,\dagger}_+\,\vert\cdots>$,
and ${\cal Q}^{1\,\dagger}_+\,{\cal Q}^{2\,\dagger}_+\,\vert\cdots>$,
where $\vert\cdots>$ is annihilated by ${\cal Q}_+^a$. However with
central charges, in the frame where only $p^+\ne 0$,
\be
<\cdots\vert\,{\cal Q}_+^1\,{\cal
Q}_-^2\,\vert\cdots>~=~iz<\cdots\vert\cdots>\ ,
\ee
and a similar argument with $1,2$ interchanged, which shows that the
states ${\cal Q}_-^a\,\vert\,0\,>$ do not vanish. Yet from 
\be
\{{\cal Q}_-^a\,,\, {\cal Q}_-^{b\,\dagger}\}~=~0\ ,
\ee
their norms must satisfy
\be
\vert\,{\cal Q}_-^a\,\vert\cdots>\,\vert^2+\vert\,{\cal 
Q}_-^{a\,\dagger}\,\vert\cdots>\,\vert^2~=~0\ ,
\ee
for $a=1,2$, implying that one is negative, or both are zero. 
This reproduces general
arguments\cite{SOHNIUS} based on  the absence of negative or zero norm
states, which show that central charges are limited to massive
representations. In the frame where only $p^+\ne 0$, the dynamic
generators of massless supersymmetry anticommute with one another. If they
do not, negative or zero-norm states are generated.   

This construction does not seem to generalize to odd dimensions. We have shown this explicitly in eleven dimensions by starting with $\vec T$ quadratic in the Grassmann numbers. There a quadratic product of a Grassmann spinor  transforms as $2$- and $3-$forms, so that to make a vector we need some c-number tensors,  either a one or two-form, but the commutation with the supercharge does not have the right form. 

On the other hand, the construction in $(9+1)$ dimensions is  straighforward. 
We consider two supercharges

$$
{\cal Q}_{+}^{(1)} ~=~ \frac{\partial}{\partial\theta} + \frac{1}{\sqrt{2}} 
p^+ \theta \ ; \quad {\cal Q}_+^{(2)} ~=~ 
\frac{\partial}{\partial \eta} + \frac{1}{\sqrt{2}} p^+ \eta \ , 
$$
where $ \theta \sim {\bf 8}_s \ , \ \ \eta \sim {\bf 8}^\prime_s $. We take the light-cone translation vector to be

$$ T^i ~=~ i\,p^+ z \theta \tgamma^i \eta ~\sim~ {\bf 8}_v\ , 
\qquad z \ {\rm real}$$
so that  this corresponds to type IIA supersymmetry. The dynamic boosts are

$$
S^{-i} ~=~ i\,z\theta \tgamma^i \eta - \frac{p^j}{p^+} \widehat S^{ij} 
- \frac{p^i}{p^+} S^{+-} \ , 
$$
with

$$\widehat S^{ij} ~=~ S^{ij} -\frac{i}{2}\left(\theta \, \tgamma^{ij}
\frac{\partial}{\partial \theta} + \eta \, \tgamma^{ij}
\frac{\partial}{\partial \eta} \right) \ , \quad {\rm and} \quad
S^{+-}~=~ -\frac{i}{2}\left(\theta \frac{\partial}{\partial\theta} + 
\eta \frac{\partial}{\partial \eta}\right) \ . 
$$
The dynamic supercharges are now 

$$ {\cal Q}^{(2)}_- ~=~ -\frac{ip^i}{\sqrt{2} \, p^+} \tgamma^i 
{\cal Q}^{(2)}_+ + i\,\sqrt{2}z \theta \ , 
$$
$$ {\cal Q}^{(1)}_- ~=~ -\frac{ip^i}{\sqrt{2} \, p^+} \tgamma^i 
{\cal Q}^{(1)}_+ - i\,\sqrt{2}z \eta \ , 
$$
The resulting supersymmetry algebra now includes central charge

$$ \{ {\cal Q}^{(1)}_+ \ , \ {\cal Q}^{(2)}_- \} ~=~  
- \{ {\cal Q}_+^{(2)} \ , \ {\cal Q}_-^{(1)} \} ~=~ i\sqrt{2}\,z \ . 
$$

We have seen that our construction leads to supersymmetry but the representations necessarily contain negative and zero-norm states. Although it is interesting to note that central charges
occur naturally whenever the light-cone translations are built out of the
Grassmann variables, the representations contain negative and zero-norm states; at best they could be used as ghost compensators of some unknown theory.

\section{Conclusion}
In this paper, we considered non-zero light-cone translations to
construct the continuous spin representations of both Poincar\'e and
Super-Poincar\'e algebra. 
We started with the Poincar\'e algebra in four space-time dimensions and
generalized it to higher dimensions. The light-cone translation vector, $\vec T$, 
is represented by spherical harmonics in $(d-2)$-dimensions for the
light-cone little group $SO(d-2)$. We find that the states can be
represented by $p^+$, the length $\Xi $ and $(d-3)$ number of angles 
of the $SO(d-2)$ spherical harmonics, and the Dynkin labels $(a_1, \cdots
, a_r)$ of  $SO(d-3)$ the short little group, with $r$ being the
rank.  There is one CSR for each representation $(a_1, \cdots , a_r)$.

If  $\vec T$ is  nilpotent, a finite number of states are 
generated by the light-cone boost, instead
of infinite number of states, but the resulting CSR contains zero or negative norm states, and 
its supersymmetric extension (in four dimensions) has a central charge. The nilpotent construction of $T^i$ in higher dimensions is not so evident. These CSRs  may be
useful in conjunction with theories where the invariances force
overcounting, as in ghost states in gauge theories.

\vspace{.2cm}
\noindent {\bf Acknowledgements}:
Three of us(AK,PR,XX) wish to acknowledge the support in part by the US Department of Energy under grant DE-FG02-97ER41029.

\vfill\eject
\end{document}